\documentclass[fleqn,10pt]{wlscirep}
\usepackage{graphicx}  
\usepackage{dcolumn}   
\usepackage{bm}        
\usepackage{amssymb}   
\usepackage{soul}

\usepackage{tikz}
\usetikzlibrary{arrows}

\usepackage{caption}
\usepackage{subcaption}
\usepackage[percent]{overpic}
\usepackage[normalem]{ulem}
\usepackage{amsmath}

\title{Large Tunable Thermophase in Superconductor -- Quantum Dot -- Superconductor Josephson Junctions}
\date{\today}

\author[1,*]{Yaakov Kleeorin}
\author[1,4]{Yigal Meir}
\author[3]{Francesco Giazotto}
\author[2,4,+]{Yonatan Dubi}

\affil[1]{Department of Physics, Ben-Gurion University of the Negev, Beer Sheva 84105, Israel}
\affil[2]{Department of Chemistry, Ben-Gurion University of the
	Negev, Beer Sheva 84105, Israel}
\affil[3]{NEST Istituto Nanoscienze-CNR and Scuola Normale Superiore, I-56127 Pisa, Italy}
\affil[4]{The Ilse Katz Institute for Nanoscale Science and Technology, Ben-Gurion University of the	Negev, Beer Sheva 84105, Israel}

\affil[*]{kleeorin@post.bgu.ac.il}
\affil[+]{jdubi@bgu.ac.il}

\begin{abstract}In spite of extended efforts, detecting thermoelectric effects in superconductors has proven to be a challenging task, due to the inherent superconducting particle-hole symmetry. Here we present a theoretical study of an experimentally attainable Superconductor -- Quantum Dot -- Superconductor (SC-QD-SC) Josephson Junction. Using Keldysh Green's functions we derive the exact thermo-phase and thermal response of the junction, and demonstrate that such a junction has highly tunable thermoelectric properties and a significant thermal response. The origin of these effects is the QD energy level placed between the SCs, which breaks particle-hole symmetry in a gradual manner, allowing, in the presence of a temperature gradient, for gate controlled appearance of a superconducting thermo-phase. This thermo-phase increases up to a maximal value of $\pm\pi/2$ after which thermovoltage is expected to develop. Our calculations are performed in realistic parameter regimes, and we suggest an experimental setup which could be used to verify our predictions.

\end{abstract}
\begin{document}
	
\flushbottom
\maketitle

\thispagestyle{empty}

\section*{Introduction}
Thermoelectric (TE) effects correspond to the response of electrical charge (via induced current or voltage) when a thermal bias
is applied across a junction. Since the warmer side has an equal excess of both particles and holes, the direction and magnitude of the TE response are determined by the asymmetry between particles and holes. Consequently, TE effects have proven to be a powerful tool in probing the density of states near the Fermi energy, particularly in materials with strong electron-electron interactions\cite{intro1,intro2,intro3}. However, in superconductors (SCs), which are a paradigmatic example of interacting electron systems, the TE response is both small in magnitude and hard to control. This is because SCs are inherently particle-hole (p-h) symmetric, and the p-h asymmetry stems primarily from impurity scattering \cite{kon,Gurevich1,Gurevich2}. Measuring a substantial and controllable TE response in SCs is therefore a major challenge.

Early experiments searching for thermocurrent in superconductors found that even the expected small thermocurrent was generally absent \cite{REV}. An explanation for the absence of thermoelectric response was proposed by Ginzburg \cite{Ginz}, who suggested, within the two fluid scheme, that the superfluid is expected, under certain conditions, to counterbalance the quasi-particle (QP) current with a non-dissipative supercurrent \cite{Gerd}. The existence of such a supercurrent is accompanied by an induced gradient in the phase of the SC order parameter  \cite{Eshel,Tinkham}.

To overcome the absence of current, experiments in which the setup comprises a bi-metallic loop (taking advantage of the fact that the SC phase has to be geometrically quantized), were proposed and performed \cite{Harlingen,Tinkham}. However, different experiments \cite{exp1} disagreed with each other and with theory \cite{theory1,Galperin}, a discrepancy which only recently may have been resolved \cite{Shelly}. Suggestions for increasing the thermal response and p-h asymmetry include using magnetic impurities \cite{Zaikin} or a ferromagnetic junction setup \cite{Giazotto}, leading to a thermo-phase of greater magnitude. However, using a magnetic field for tuning the system parameters  \cite{Tero} leads to substantial experimental limitations.

\begin{figure}[ht]
	
	\includegraphics[width=10cm]{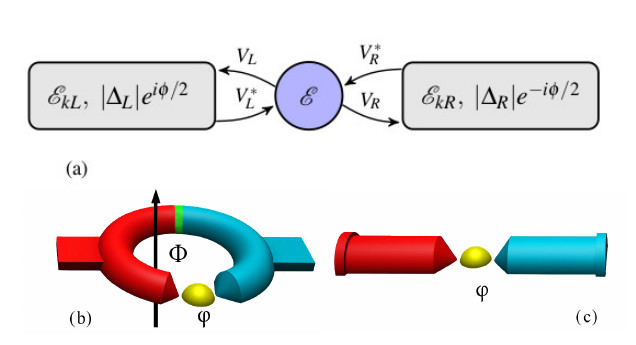}

	\caption{(a) Schematic representation of the SC-QD-SC setup. The two SC leads are characterized by their spectrum, gap energies and their phases. The QD contains a single degenerate level of energy $\epsilon$. (b) Closed loop experimental setup. In this setup the phase difference $\Delta\phi$ is geometrically constrained, so it cannot always compensate for the thermal quasi-particle current. (c) Open experimental setup, where $\Delta\phi$ can assume arbitrary values.}
	\label{fig:setup}
	
\end{figure}

In spite of all these efforts, the challenge of devising a SC system which exhibits substantial TE effects and with a large degree of control is yet to be met. Here, we demonstrate that in a SC-quantum dot (QD)-SC setup (schematically depicted in Fig.~\ref{fig:setup}(a)), the TE response can be considerably larger than in SC tunnel junctions \cite{Eshel}, where measurable thermo-phase can only arise around the transition temperature. Control over its magnitude can be achieved by a gate voltage, which shifts the energy levels of the QD, allowing for breaking of the p-h symmetry even for ideal SC electrodes, thus enabling experimental control of the magnitude and direction of the thermal response. It is important to note, that such a setup is within current experimental capabilities \cite{expcap1,expcap2,expcap3,expcap4}, making our predictions experimentally verifiable.

\section*{Model} Our model consists of bulk s-type superconductors as leads, with individual gap energies and arbitrary phases (taken symmetrically for convenience),  and a single QD level in between. The Hamiltonian for the SC-QD-SC junction is given by $H=H_L+H_R+H_{QD}+H_V$, with the lead Hamiltonians $H_s$ ($s=L,R$) given by
\begin{equation}
H_s= \sum_{s,k,\sigma} \epsilon_{ks} c_{sk\sigma}^\dagger c_{sk\sigma} +\sum_{s,k,\sigma}  \Delta_s c_{sk\sigma}^\dagger c_{s-k-\sigma}^\dagger+H.c
\end{equation}
where $c_{sk\sigma}^\dagger (c_{sk\sigma})$ is the creation (annihilation) of an electron on side $s$ with momentum $k$, spin $\sigma$. The order parameter is complex, $\Delta_s=e^{i \phi_s}|\Delta_s|$ and the phase difference is taken, without loss of generality,  as $\phi_L=-\phi_R=\phi /2$. The chemical potential in the SC leads is defined as the zero of energy. We first start with a non-interacting, single-level QD. In this case, the QD Hamiltonian $H_{QD}$  and the hopping Hamiltonian between the QD and the SCs, $H_{V}$, are
\begin{equation}
H_{QD}= \sum_{\sigma} \epsilon_\sigma d_{\sigma}^\dagger d_{\sigma}~, \quad H_{V}= \sum_{s,k,\sigma} V_{ks} c_{sk\sigma}^\dagger d_{\sigma}+H.c
\end{equation}
where $d_{\sigma}^\dagger (d_{\sigma})$ is the creation (annihilation) of an electron on the dot with spin $\sigma$. While the calculation is quite general, in the present context we assume spin degeneracy,  $\epsilon_\uparrow = \epsilon_\downarrow \equiv \epsilon$, and uniform tunneling $V_{ks} \equiv V_s$. From this Hamiltonian, the currents and other quantities are calculated using the non-equilibrium Green's function method, as described in the Methods section.

\section*{Results}
We start by addressing the general form of the current. Substituting the Green's function (Eq.~\ref{Gfunc} in the Methods section) into the expression for the current, we find that the current can be generally divided into three terms: quasi-particle current, $I_{QP}$, Josephson (pair) current, $I_{sc}$,  and a term involving pair-QP transition, $I_{pair-QP}$,
\begin{eqnarray}
I&=&I_{QP}(\epsilon,\phi,T,\Delta T)+ I_{pair-QP}(\epsilon,\phi,T,\Delta T)\cos^2(\phi/2)
\nonumber\\
 &+& I_{sc}(\epsilon,\phi,T,\Delta T)\sin(\phi)
\label{organize1}
\end{eqnarray}

A temperature dependence exists in all the terms through the temperature dependence of the superconducting order parameter. The usefulness of this form for the current stems from the fact that within the relevant parameter range, discussed here,
the phase dependence of the amplitude of the various current terms is negligible. This phase dependence originates from multiple reflections between the QD and the leads, giving rise to higher harmonic processes with a non trivial phase function. These reflections diminish as a function of $\Gamma_s/\epsilon$, i.e. as the energy level in the QD moves away from the Fermi level, and as a result Cooper pairs have smaller probability of tunneling across the junction. In the  parameter range for which the thermo-phase is appreciable -- the tunnel junction regime -- the ratio $\Gamma_s/\epsilon$ is small and thus multiple reflections can practically be neglected.

\begin{figure}[ht]
	
	\includegraphics[width=10cm]{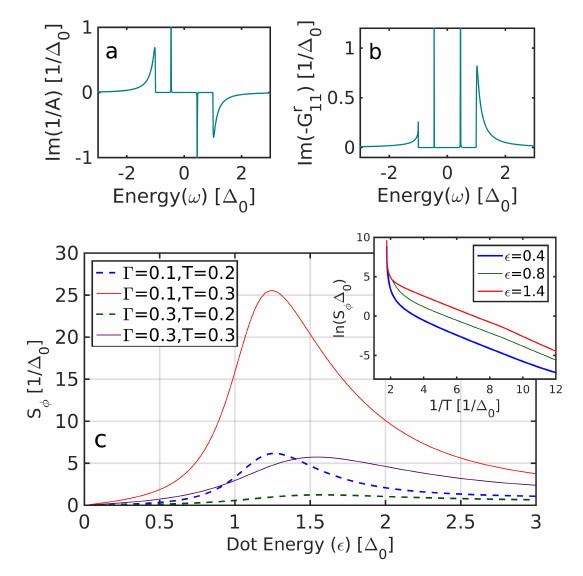}
	
	\caption{{\bf{(a): }}Imaginary part of 1/A, the pair channel of transmission through the dot, as a function of energy, for $\Gamma=0.1, \epsilon=0.5, \phi=\pi/2$. The narrow peaks are the ABS inside the gap {\bf{(b): }}Imaginary part of $-G_{11}^r$, the quasi-particle channel of transmission through the dot, which is proportional to the DOS of electrons on the dot, as a function of energy, for the same parameters as (a). The p-h asymmetry is visible in the continuum. {\bf{(c): }}The thermo-phase Seebeck coefficient $S_{\phi}$ (Eq.~\ref{TPSC}) as a function of dot energy, for various $T$ and $\Gamma$. We can see that the peak position depends only on $\Gamma$ and $\Delta$ as will be discussed in the text. \bf{Inset: }\normalfont $\ln(S_{\phi})$ as a function of inverse temperature for various dot levels, $\epsilon=0.4,0.8,1.4$.}
	\label{fig:seebeck}
\end{figure}



In an open junction setup (Fig.~\ref{fig:setup}(c)), with no externally imposed constrains over the thermo-phase, the thermally induced current is completely canceled by the appearance of a thermo-phase across the junction \cite{Ginz}. This serves as the definition for the thermo-phase $\phi_{th}$:
\begin{equation}
\label{condition}
I(\epsilon, T, \Delta T, \phi_{th})=0
\end{equation}
\subsection*{Linear Response}
In the linear response regime (linear in $\Delta T$), assuming a symmetric junction, one can write the different terms in Eq.~\ref{organize1} explicitly:
\begin{align}
&I_{QP}=\frac{e}{h}\sum_\sigma \int_{-\infty}^\infty d\omega \frac{df(\omega)}{dT}\Gamma\ {\rm Re}[\rho(\omega)]\ {\rm Im}[-G^r_{11}(\omega)]\Delta T
\label{organize2}
\\
&I_{sc}=\frac{e}{h}\sum_\sigma \int_{-\infty}^\infty d\omega f(\omega)\frac{\Delta^2\Gamma^2}{\omega^2-\Delta^2}\ {\rm Im}[1/A(\omega)]
\end{align}
where the parameters $\Gamma$ and $\Delta$ were taken equal on both sides of the junction. In these equations, the expression for $G_{11}^r$ and $A$, whose contributions to the current are plotted in Fig.~\ref{fig:seebeck}, is given by $A(\omega)=\mathrm{Det}[(\hat{g}_0^r(\omega))^{-1}-\hat{\Sigma}^r(\omega)]$(Fig.~\ref{fig:seebeck}(a)) and $G^r_{11}(\omega)=(\omega+\epsilon-\Sigma^r_{11})/A(\omega)$(Fig.~\ref{fig:seebeck}(b)). The transmission channel for the QPs, $Im[-G_{11}^r(\omega)]$, demonstrates the asymmetry of transmission as a function of energy, required to generate a thermoelectric response. On the other hand, the transmission channel for pairs, $1/A(\omega)$, is driven by  a superconducting phase difference, generated to compensate for the QP contribution, and thus  does not require p-h asymmetry. As can be seen in Fig.~\ref{fig:seebeck}(a,b), both the QP channel and the pair channel contain sharp resonances which are Andreev bound states (ABS) (though the ABS do not participate in the QP transport due to $Re[\rho(\omega)]$ term in Eq.~\ref{organize2}, as $\rho$ has a real part only outside the gap). The pair-QP transition term $I_{pair-QP}$ vanishes identically, since this thermal transport process is perfectly particle-hole symmetric (mathematically, writing $I_{pair-QP}$ as an integral similar to Eq.~\ref{organize2}, the integrand is an odd function of $\omega$, as a result of a symmetric transmission channel in this process).

Since in linear response one can define $I=\sigma \Delta\phi +S_{\phi}\sigma \Delta T$, in analogy to the Seebeck coefficient, we can define the thermo-phase Seebeck coefficient (TPSC) in a similar manner,
\begin{equation}
 S_{\phi}=-\Big(\frac{\Delta \phi}{\Delta T}\Big)_{I=0}=\frac{dI_{QP}/d\Delta T}{I_{sc}}
\label{TPSC} \end{equation}

In Fig.~\ref{fig:seebeck}(c), the TPSC $S_{\phi}$ is plotted as a function of dot level energy $\epsilon$, for various temperatures $T$ and couplings $\Gamma$. $S_{\phi}$ consistently peaks around $|\epsilon|=\Delta+a\Gamma$, with the factor $a$ being typically $a\sim 1-2.5$ for the relevant parameters ($\Gamma>0.05$). The TPSC peak occurs when the dot energy is slightly above the SC coherence peaks in the BCS DOS, at the point that maximizes the interplay of p-h asymmetry and transmission. This is similar in nature to the Seebeck coefficient peak through a QD between normal leads \cite{Yoni}, which resides a distance $~\Gamma$ above (or below) the QD energy level resonance. In the inset of Fig.~\ref{fig:seebeck}(c) we plot the inverse temperature dependence of the TPSC on a log scale, for various level energies. For $T\ll\Delta$ the leading contribution to the temperature dependence of the TPSC stems from the activated form of the the Fermi function in the QP term (\ref{organize2}), which can be approximated by $\sim e^{-E_g/T}$, where $E_g$ is an activation energy. Indeed, the logarithmic slopes of $S_{\phi}$, depicted in the inset of Fig.~\ref{fig:seebeck}(c), are  linear with an activation gap $\Delta_0$, as expected for QPs. This behavior works rather well for most of the relevant temperature range. In the opposite limit, for T approaching the SC transition temperature $T_c$, the TPSC diverges due to vanishing of the Josephson term, as $1/\Delta^2\sim(T-T_c)^{-1}$.

\subsection*{Beyond Linear Response}
\begin{figure}[ht]
	\includegraphics[width=10cm ]{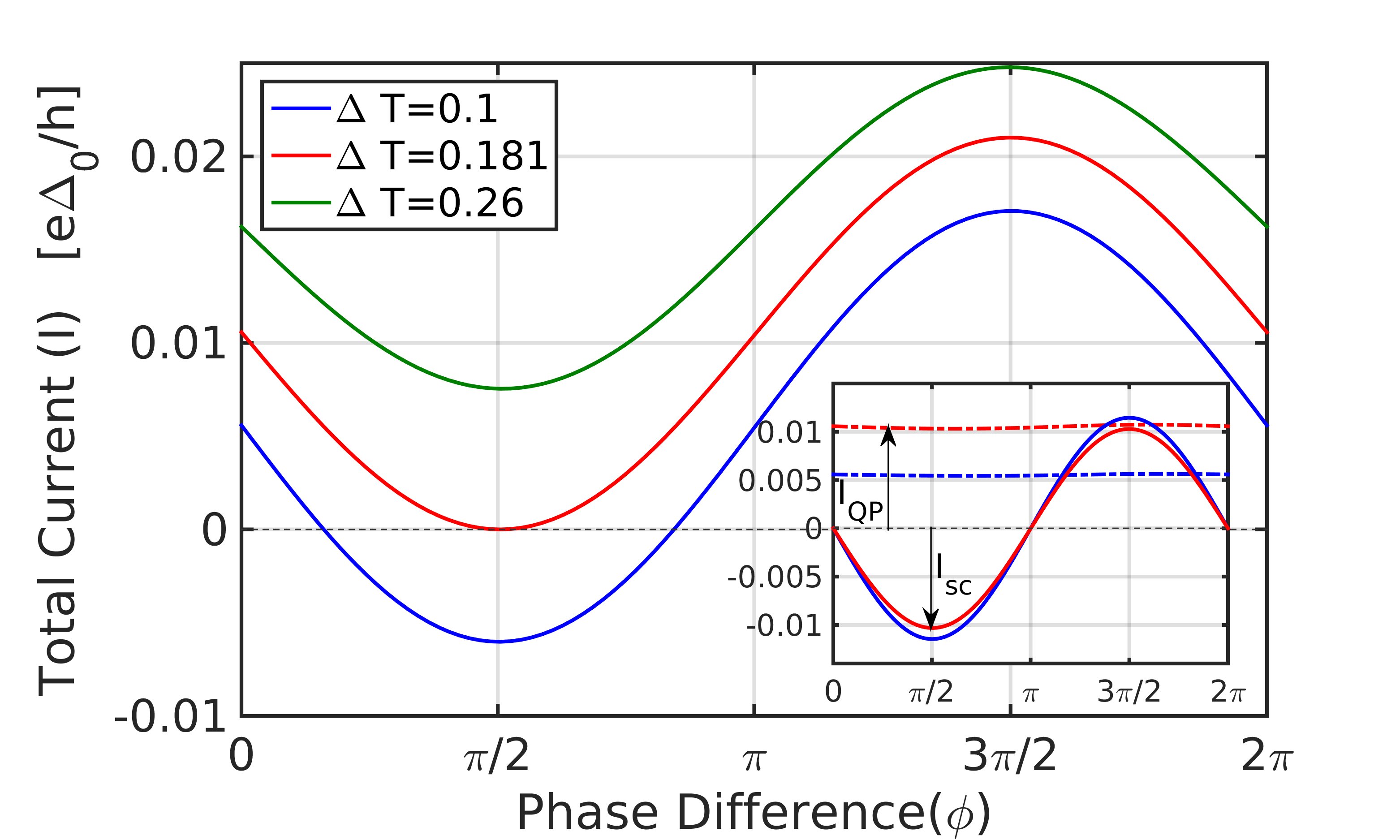}
	\caption{The total current $I$ as a function of phase difference for various temperature differences $\Delta T$, for  $\epsilon=1.1,  T=0.2, \Gamma=0.1$.{\bf{ Inset:}} current divided into the two contributions: quasi-particle current and Josephson current. The first (QP) term in equation (3) gives the up shift in current due to temperature bias and the third (Josephson) term gives the amplitude of the modulation with phase. Phase dependence is negligible in the QP term (dot-dashed line).}
	\label{fig:plots1}
\end{figure}

The formulation described in the previous section applies, in fact, also beyond the linear response regime in $\Delta T$, where the main deviation from linear response stems from the difference in order parameters on both sides due to thermal difference. The full analytical expression, including all contributions, is quite long and thus will not be shown here. Fig.~\ref{fig:plots1} depicts the total current as a function of phase for several values of temperature difference, $\Delta T$. The general division of the current into the three terms (Eq.~\ref{organize1}) holds also beyond the linear response regime, as can be seen in the inset of Fig.~\ref{fig:plots1}, which shows the QP and the Josephson contributions to the total current  (the contribution from the pair-QP transition term $I_{pair-QP}$ still vanishes). The Josephson term is modified due the difference in $\Delta$ between the two sides  \cite{AndreevReflection}. As can be clearly seen from the figure, the QP term is almost insensitive to phase difference, but sensitive to changes in temperature difference, while the Josephson term oscillates with the phase difference, but weakly sensitive to temperatures far from the SC transition temperature.

 As the temperature difference increases beyond a critical value $\Delta T_c$ (the red curve in Fig.~\ref{fig:plots1}, corresponding to $\Delta T=0.181$ for the depicted set of parameters) the QP current reaches a value such that the Josephson current can no longer compensate for it (for $\Delta T=\Delta T_c$ the thermo-phase is exactly $\pm\pi/2$). If the total current is kept at zero, an effective voltage will develop in this regime, which will give rise to a time-dependent AC response (as in the AC Josephson effect), an effect which has in fact been measured in tunnel junctions \cite{voltage}. We leave the time-dependent thermal Josephson effect for a future study, and concentrate here on $\Delta T$  below the critical value $\Delta T_c$.

\begin{figure}[ht]
	\includegraphics[width=10cm ]{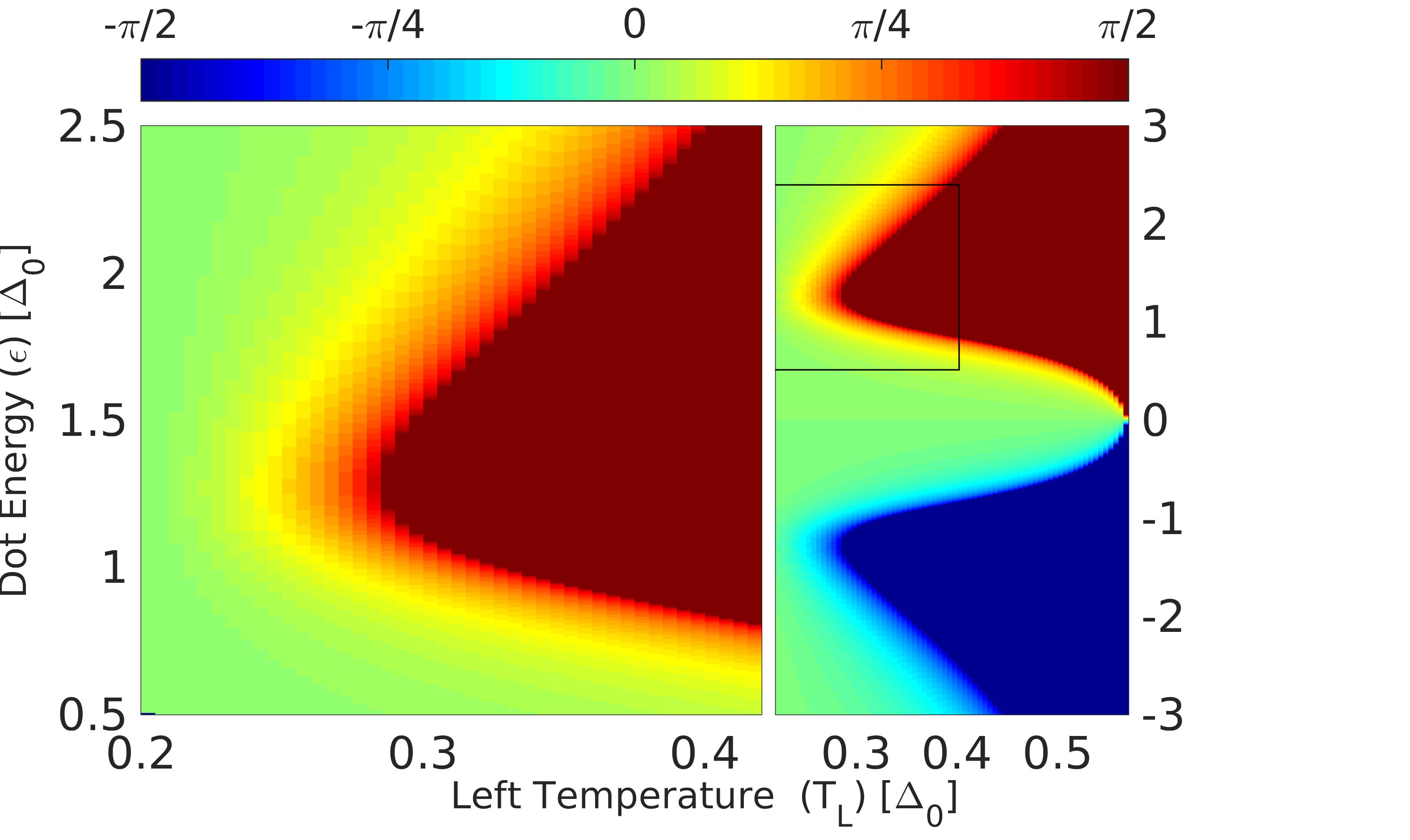}
	\caption{{\bf{Left panel:}} The thermo-phase $\phi_{th}$ as a function of dot energy and $T_L$, for $T_R=0.2, \Gamma=0.1$. The $\pm\pi/2$ plateau (red or blue) means that the quasi-particle current ($I_{QP}$) has reached or exceeded the Josephson amplitude ($I_{c}$). {\bf{Right panel:}} the same plot for a larger range of parameters, including negative dot energies. The thermo-phase is odd with respect to $\epsilon$.}
	\label{fig:surf}
\end{figure}

 Solving the condition (\ref{condition}) for vanishing current, we plot in Fig.~\ref{fig:surf} the thermo-phase $\phi_{th}$ as a function of the left lead temperature $T_L$ (for fixed $T_R$) and QD level energy $\epsilon$. The region of $\phi_{th}=\pm\pi/2$ (red or blue plateau in Fig.~\ref{fig:surf}) corresponds to the regime for which $\Delta T\geq \Delta T_c$, and is not covered in this work. The value of the critical temperature difference as a function of dot energy can be read from Fig.~\ref{fig:surf} as the contour of the $\pm\pi/2$ plateau. The value of the critical $\Delta T_c$ can be directly measured in experiments, by applying a temperature difference and monitoring for which $\Delta T$ a finite current (or voltage) begins to appear.

\subsection*{Coulomb Interaction}

\begin{figure}[ht]
	\includegraphics[width=12cm]{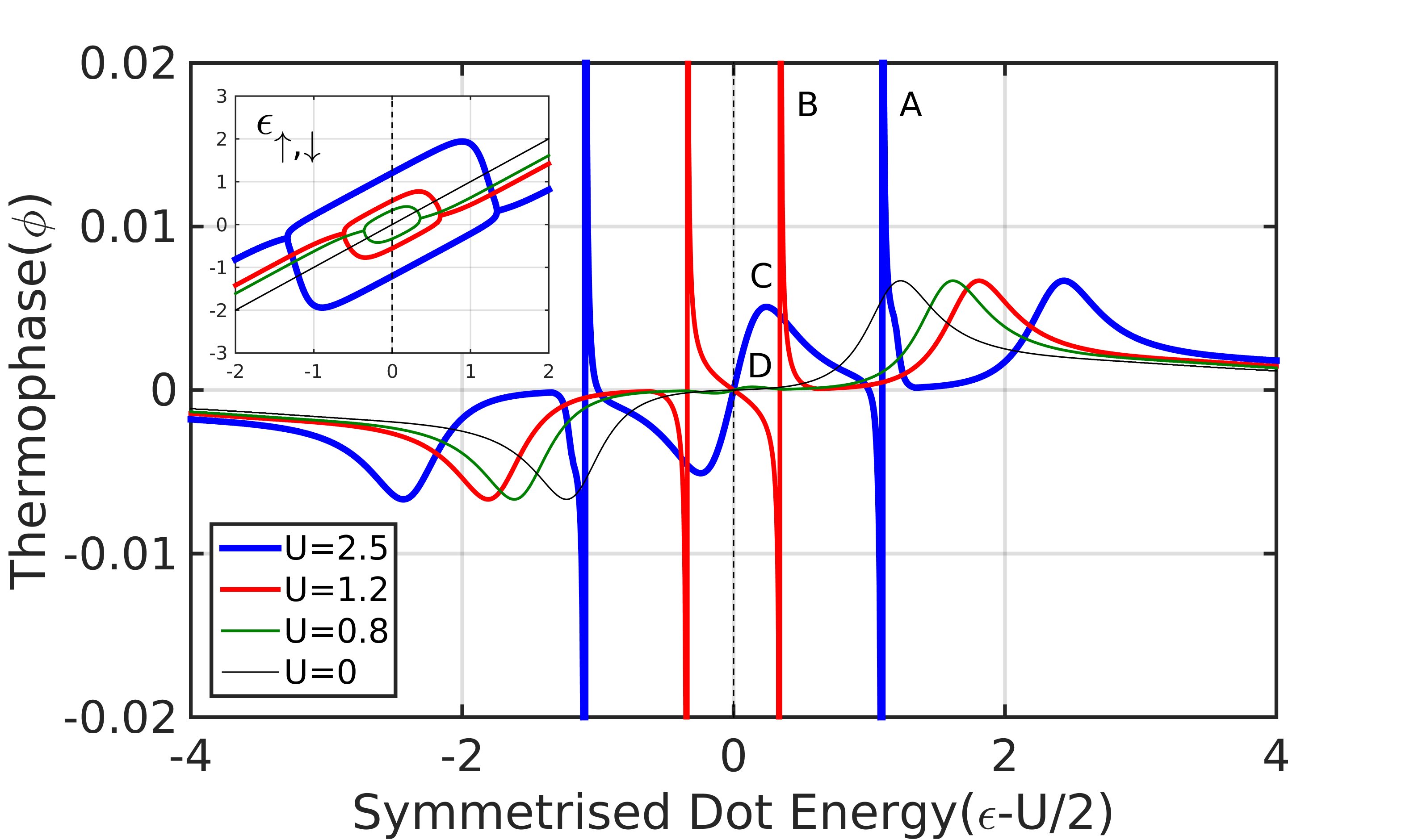}
	\caption{Thermo-phase as a function of symmetrized bare dot energy for various values of interaction strength U. $T=0.1$, $\Delta T=0.05$, $\Gamma=0.1$ {\bf Inset:} renormalized dot energies as a function of symmetrized bare dot energy. A separation between the renormalized dot energies appears in the doublet regime (singly occupied), where a magnetic moment is formed.}
	\label{fig:interaction}
\end{figure}

So far we have ignored the on-site interaction on the QD, which may be important, for example, in the Coulomb blockade regime \cite{Anderson}. In order to address this, we add to the Hamiltonian an on-site Coulomb interaction, represented by a term $H_U=U n_\uparrow n_\downarrow$, where  $n_{\sigma}\equiv d_{\sigma}^\dagger d_{\sigma}$.  Within the Hartree-Fock (HF) approximation, the dot levels are renormalized according to $\epsilon_{\sigma}=\epsilon + U \langle n_{\overline{\sigma}}\rangle$, where the dot occupations $\langle n_{\sigma}\rangle$ are calculated self-consistently ($\overline{\sigma}$ is the spin opposite to $\sigma$). Once the dot levels and occupations are determined (inset in Fig.~\ref{fig:interaction}), the thermo-phase can be calculated using Eq.~\ref{condition}.


Compared to the non-interacting problem, the interaction introduces a new regime where the dot is singly occupied, the effective spin energy levels split \cite{ABS}, and a magnetic moment forms.  This splitting suppresses the pair tunneling amplitude, where eventually the Josephson term becomes smaller and, unless smeared by temperature, changes sign, leading to a $\pi$-junction transition \cite{arovas,Zhu}. It is important to note that the HF approximation, while found to generally describe the SC-QD-SC physics very well \cite{arovas, rodero}, will not be valid in the Kondo regime, where $T_k>\Delta$. In addition, it may give qualitatively inaccurate values for the boundaries of the singly occupied regime \cite{zonda}. These failures of the HF approximation are rather limited in the small $\Gamma/\Delta$ limit \cite{zonda}, which is the regime of interest in the present work, and therefore we can safely proceed with its usage. Furthermore, we note that although there seems to be an apparent spin symmetry breaking from the form of the HF solution, these spin-asymmetric solutions are doubly-degenerate with the degenerate solutions having opposite spins. It is thus important to take both solutions into account to preserve spin symmetry. In Fig.~\ref{fig:interaction} we show the thermo-phase $\phi_{th}$ as a function of the bare dot energy (shifted by half the Coulomb interaction), for various values of $U$. We first note that the thermo-phase is symmetric not around $\epsilon=0$, but around the new (and only) point of particle-hole symmetry, $\epsilon-U/2=0$. There are other points, inside the singly occupied region, where the QP term (and consequently also the thermo-phase) vanish, but this is due to cancellation of contributions and not because of p-h symmetry.  Inside this region, we also see sharp $\pm\pi/2$ peaks (points A,B in Fig.~\ref{fig:interaction}) which correspond to a vanishing Josephson term during the $\pi$-junction transition, where any thermal gradient will produce the maximal thermo-phase of $\pm \pi/2$. Other new features (such as a small peak at point C and a tiny peak at point D in Fig.~\ref{fig:interaction}) emerge from the non-monotonous behavior of the 
QP term, in the singly occupied regime. In this regime, the contributions from the two spin levels have opposite signs and their magnitude difference also changes sign as a function of average dot energy. The features outside this region, however, are unaffected by the interaction except for the trivial shift away from the p-h symmetry point by U/2.

\section*{Discussion}
All the results presented in this paper can be directly tested experimentally. To measure the thermoelectric effect and the thermo-phase, we suggest the experimental setup depicted in Fig.~\ref{fig:setup}(b). It consists of a SC ring with one branch including a QD while the other branch including a thin insulating barrier. One side of the ring is heated in order to create a temperature gradient, and as a result, a unidirectional circulating thermocurrent arises. This setup makes use of the geometrical constraint on the gauge invariant phase, and of the fact that the phase drop occurs primarily at the point of most resistance \cite{Ambegauker} (which is the QD as opposed to the insulating barrier). This implies that the phase difference across the QD junction is $\phi=2 \pi (\Phi/\Phi_0+n)$, where $\Phi$ is the magnetic flux penetrating the ring and $\Phi_0=hc/2e$ is the flux quantum. Since there is no external magnetic flux, the phase difference across the QD, necessary to produce the supercurrent that cancels the thermal current in the bulk of the SC, is accompanied by a generation of a magnetic flux through the ring. This flux, in fact, arises from supercurrents running on the surface of the ring\cite{Harlingen}. This experimentally measurable flux can be continuously modified by the applied temperature gradient, or the position of the dot level energy. In order to measure the critical temperature difference, an open setup (Fig.~\ref{fig:setup}(c)) can also be utilized (where no phase detection is necessary). The temperature difference for which effective thermovoltage begins to appear, is the critical temperature difference. Ref.\cite{expcap3} has already applied setups that involve SCs and a QD, while Ref. \cite{expcap2} has already demonstrated  applying a temperature bias in SC\cite{expcap2}, but these two approaches have yet to be experimentally explored together.

In summary, we have demonstrated that a superconductor - quantum dot - superconductor junction can serve as a model system to study thermoelectric effects in SC systems, as it exhibits a large and controllable TE response. The current response to a temperature difference has been studied as a function of the most important control parameters, namely temperature, gate voltage and dot-electrode couplings. Specific experimental realizations to test our predictions have been suggested, and we believe that they are well within current experimental capabilities. Further studies that examine the AC thermal Josephson effect (beyond the critical temperature difference) are currently under way.

\section*{Methods}

In order to find the current across the junction we calculate the Green's function in Nambu space \cite{Zhu}, $\hat{G}_{\sigma}^r(t)=-i\theta(t)\langle \{ \Psi (t), \Psi^\dagger(0)\} \rangle$, where $\Psi^\dagger=\begin{pmatrix}
d_{\sigma}^\dagger \\ d_{\overline{\sigma}}\end{pmatrix} $ is the Nambu particle-hole spinor.
We find the relevant self energies using the equations of motion, in Nambu space:
\begin{equation}
\hat{\Sigma}^r_{s}(\omega)=-\frac{i}{2}\Gamma_{s}\rho_s(\omega)\begin{pmatrix}
1 & -\frac{\Delta}{\omega}e^{i\phi_s}
\\
-\frac{\Delta}{\omega}e^{-i\phi_s} & 1
\end{pmatrix}
\end{equation}
\begin{equation}
\rho_s(\omega)=\begin{cases}
\frac{|\omega|}{\sqrt{\omega^2-\Delta_s^2}} & |\omega|>\Delta_s \\
\frac{\omega}{i\sqrt{\Delta_s^2-\omega^2}} & |\omega|<\Delta_s
\end{cases}
\end{equation}

where $\Gamma_s=2\pi V_s^2 N_s(0)$, $N_s(0)$ being the normal metal density of states (DOS). $\rho_s$ can be regarded as the generalized DOS in the superconductor, normalized by the normal metal value, where there is an imaginary ($|\omega|<\Delta_s$) contribution from inside the gap. Applying the self energies to the Dyson equation \cite{Zhu} we find the retarded Green's Function in Fourier space $\hat{G}_{\sigma}^r(\omega)=[(\hat{g}_0^r(\omega))^{-1}-\hat{\Sigma}^r(\omega)]^{-1}$, where
\begin{equation}
\hat{g}^r_0(\omega)^{-1}=\begin{pmatrix}
\omega-\epsilon+i 0^{+} & 0
\\
0 & \omega+\epsilon+i 0^{+}
\end{pmatrix}
\label{Gfunc} \end{equation}
and $\Sigma^r=\Sigma^r_L+\Sigma^r_R$. Using $J_{L(R)}=-e\dot{n}_{L(R)}$, we express the current in terms of the Green's functions on the dot \cite{Meir}, $
J=-\frac{e}{h}\sum_\sigma\int d\omega \ {\rm Re}[(\hat\Sigma_L(\omega)-\hat\Sigma_R(\omega))\hat{G}_\sigma(\omega)]^<_{11}$,
where the lesser Green's functions $G_\sigma(t)^<_{ij}\equiv i<\Psi_i^\dagger \Psi_j(t)>$, and we can calculate the term in the square brackets  in the expression for $J$ using the Langreth relation \cite{Langreth} $ [A(\omega)B(\omega)]^<=A^r(\omega)B^<(\omega)+A^<(\omega)B^a(\omega)$.

From a numerical perspective, a broadening of the superconducting gap energy is required to avoid divergence of the superconducting DOS. A suitable Dynes Broadening \cite{Dyne1,Dyne2} is required, and if done carefully (the broadening should be mutually conjugate for particles and holes, namely $\Delta(\omega)=\Delta_0-i \ \text{sign}(\omega)\eta $), it enables one to directly see the contribution from the Andreev bound states \cite{Zhu,ABS}, which are usually numerically elusive (being ideally a delta function contribution to the local DOS), as can be seen in Fig.~\ref{fig:seebeck}(a),~\ref{fig:seebeck}(b). In all the calculations the zero temperature SC order parameter on both sides was set as the unit energy, $\Delta_s(T=0)\equiv\Delta_0=1$, and all other energy values are measured in units of $\Delta_0$. The value of the Dynes broadening parameter used in our calculations is $\eta=10^{-4}$, but the results are largely independent of this value.


\begin{thebibliography}{10}
	\expandafter\ifx\csname url\endcsname\relax
	\def\url#1{\texttt{#1}}\fi
	\expandafter\ifx\csname urlprefix\endcsname\relax\def\urlprefix{URL }\fi
	\providecommand{\bibinfo}[2]{#2}
	\providecommand{\eprint}[2][]{\url{#2}}
	
	\bibitem{intro1}
	\bibinfo{author}{Behnia, K.}, \bibinfo{author}{Jaccard, D.} \&
	\bibinfo{author}{Flouquet, J.}
	\newblock \bibinfo{title}{On the thermoelectricity of correlated electrons in
		the zero-temperature limit}.
	\newblock \emph{\bibinfo{journal}{Journal of Physics: Condensed Matter}}
	\textbf{\bibinfo{volume}{16}}, \bibinfo{pages}{5187} (\bibinfo{year}{2004}).
	
	\bibitem{intro2}
	\bibinfo{author}{Kim, K.-S.} \& \bibinfo{author}{P\'epin, C.}
	\newblock \bibinfo{title}{Thermopower as a signature of quantum criticality in
		heavy fermions}.
	\newblock \emph{\bibinfo{journal}{Phys. Rev. B}} \textbf{\bibinfo{volume}{81}},
	\bibinfo{pages}{205108} (\bibinfo{year}{2010}).
	
	\bibitem{intro3}
	\bibinfo{author}{Zlati\ifmmode~\acute{c}\else \'{c}\fi{}, V.},
	\bibinfo{author}{Monnier, R.}, \bibinfo{author}{Freericks, J.~K.} \&
	\bibinfo{author}{Becker, K.~W.}
	\newblock \bibinfo{title}{Relationship between the thermopower and entropy of
		strongly correlated electron systems}.
	\newblock \emph{\bibinfo{journal}{Phys. Rev. B}} \textbf{\bibinfo{volume}{76}},
	\bibinfo{pages}{085122} (\bibinfo{year}{2007}).
	
	\bibitem{kon}
	\bibinfo{author}{Kon, L.}
	\newblock \bibinfo{title}{Thermoelectric effect in superconductors with
		nonmagnetic localized states}.
	\newblock \emph{\bibinfo{journal}{Zhurnal Eksperimental'noi i Teoreticheskoi
			Fiziki}} \textbf{\bibinfo{volume}{70}}, \bibinfo{pages}{286--91}
	(\bibinfo{year}{1976}).
	
	\bibitem{Gurevich1}
	\bibinfo{author}{Gal'Perin, Y.~M.}, \bibinfo{author}{Gurevich, V.} \&
	\bibinfo{author}{Kozub, V.}
	\newblock \bibinfo{title}{Acoustoelectric and thermoelectric effects in
		superconductors}.
	\newblock \emph{\bibinfo{journal}{Soviet Journal of Experimental and
			Theoretical Physics Letters}} \textbf{\bibinfo{volume}{17}},
	\bibinfo{pages}{476} (\bibinfo{year}{1973}).
	
	\bibitem{Gurevich2}
	\bibinfo{author}{Gal'Perin, Y.~M.}, \bibinfo{author}{Gurevich, V.} \&
	\bibinfo{author}{Kozub, V.}
	\newblock \bibinfo{title}{Nonlinear acoustic effects in superconductors}.
	\newblock \emph{\bibinfo{journal}{Soviet Journal of Experimental and
			Theoretical Physics}} \textbf{\bibinfo{volume}{38}}, \bibinfo{pages}{517}
	(\bibinfo{year}{1974}).
	
	\bibitem{REV}
	\bibinfo{author}{Falco, C.~M.} \& \bibinfo{author}{Garland, J.~C.}
	\newblock \bibinfo{title}{Thermoelectric effects in superconductors}.
	\newblock In \emph{\bibinfo{booktitle}{Nonequilibrium Superconductivity,
			Phonons, and Kapitza Boundaries}}, \bibinfo{pages}{521--540}
	(\bibinfo{publisher}{Springer}, \bibinfo{year}{1981}).
	
	\bibitem{Ginz}
	\bibinfo{author}{Ginzburg, V.}
	\newblock \bibinfo{title}{On thermoelectric phenomena in superconductors}.
	\newblock \emph{\bibinfo{journal}{J. Phys. USSR}} \textbf{\bibinfo{volume}{8}},
	\bibinfo{pages}{148--156} (\bibinfo{year}{1944}).
	
	\bibitem{Gerd}
	\bibinfo{author}{Schon, G.}
	\newblock \bibinfo{title}{Thermoelectric effects in superconductors}.
	\newblock In \emph{\bibinfo{booktitle}{Festkoperprobleme (Advances in Solid
			State Physics}}, vol.~\bibinfo{volume}{21}, \bibinfo{pages}{341}
	(\bibinfo{publisher}{Vieweg, Braunschweig}, \bibinfo{year}{1981}).
	
	\bibitem{Eshel}
	\bibinfo{author}{Guttman, G.~D.}, \bibinfo{author}{Nathanson, B.},
	\bibinfo{author}{Ben-Jacob, E.} \& \bibinfo{author}{Bergman, D.~J.}
	\newblock \bibinfo{title}{Thermoelectric and thermophase effects in josephson
		junctions}.
	\newblock \emph{\bibinfo{journal}{Phys. Rev. B}} \textbf{\bibinfo{volume}{55}},
	\bibinfo{pages}{12691--12700} (\bibinfo{year}{1997}).
	
	\bibitem{Tinkham}
	\bibinfo{author}{Smith, A.~D.}, \bibinfo{author}{Tinkham, M.} \&
	\bibinfo{author}{Skocpol, W.~J.}
	\newblock \bibinfo{title}{New thermoelectric effect in tunnel junctions}.
	\newblock \emph{\bibinfo{journal}{Phys. Rev. B}} \textbf{\bibinfo{volume}{22}},
	\bibinfo{pages}{4346--4354} (\bibinfo{year}{1980}).
	
	\bibitem{Harlingen}
	\bibinfo{author}{Garland, J.} \& \bibinfo{author}{Harlingen, D.~V.}
	\newblock \bibinfo{title}{Thermoelectric generation of flux in a bimetallic
		superconducting ring}.
	\newblock \emph{\bibinfo{journal}{Physics Letters A}}
	\textbf{\bibinfo{volume}{47}}, \bibinfo{pages}{423 -- 424}
	(\bibinfo{year}{1974}).
	
	\bibitem{exp1}
	\bibinfo{author}{Zavaritskii, N.~V.}
	\newblock \bibinfo{title}{Observation of superconducting current excited in a
		superoonducior by heat flow}.
	\newblock \emph{\bibinfo{journal}{Soviet Journal of Experimental and
			Theoretical Physics Letters}} \textbf{\bibinfo{volume}{19}},
	\bibinfo{pages}{126–128} (\bibinfo{year}{1974}).
	
	\bibitem{theory1}
	\bibinfo{author}{S.N.~Artemenko, A.~V.}
	\newblock \bibinfo{title}{The thermoelectric field in superconductors}.
	\newblock \emph{\bibinfo{journal}{Soviet Journal of Experimental and
			Theoretical Physics}} \textbf{\bibinfo{volume}{43}}, \bibinfo{pages}{548}
	(\bibinfo{year}{1976}).
	
	\bibitem{Galperin}
	\bibinfo{author}{Galperin, Y.~M.}, \bibinfo{author}{Gurevich, V.~L.},
	\bibinfo{author}{Kozub, V.~I.} \& \bibinfo{author}{Shelankov, A.~L.}
	\newblock \bibinfo{title}{Theory of thermoelectric phenomena in
		superconductors}.
	\newblock \emph{\bibinfo{journal}{Phys. Rev. B}} \textbf{\bibinfo{volume}{65}},
	\bibinfo{pages}{064531} (\bibinfo{year}{2002}).
	
	\bibitem{Shelly}
	\bibinfo{author}{{Shelly}, C.~D.}, \bibinfo{author}{{Matrozova}, E.~A.} \&
	\bibinfo{author}{{Petrashov}, V.~T.}
	\newblock \bibinfo{title}{{Resolving thermoelectric paradox in
			superconductors}}.
	\newblock \emph{\bibinfo{journal}{ArXiv e-prints}}  (\bibinfo{year}{2015}).
	\newblock \eprint{1508.07249}.
	
	\bibitem{Zaikin}
	\bibinfo{author}{Kalenkov, M.~S.}, \bibinfo{author}{Zaikin, A.~D.} \&
	\bibinfo{author}{Kuzmin, L.~S.}
	\newblock \bibinfo{title}{Theory of a large thermoelectric effect in
		superconductors doped with magnetic impurities}.
	\newblock \emph{\bibinfo{journal}{Phys. Rev. Lett.}}
	\textbf{\bibinfo{volume}{109}}, \bibinfo{pages}{147004}
	(\bibinfo{year}{2012}).
	
	\bibitem{Giazotto}
	\bibinfo{author}{Giazotto, F.}, \bibinfo{author}{Heikkil\"a, T.~T.} \&
	\bibinfo{author}{Bergeret, F.~S.}
	\newblock \bibinfo{title}{Very large thermophase in ferromagnetic josephson
		junctions}.
	\newblock \emph{\bibinfo{journal}{Phys. Rev. Lett.}}
	\textbf{\bibinfo{volume}{114}}, \bibinfo{pages}{067001}
	(\bibinfo{year}{2015}).
	
	\bibitem{Tero}
	\bibinfo{author}{Ozaeta, A.}, \bibinfo{author}{Virtanen, P.},
	\bibinfo{author}{Bergeret, F.~S.} \& \bibinfo{author}{Heikkil\"a, T.~T.}
	\newblock \bibinfo{title}{Predicted very large thermoelectric effect in
		ferromagnet-superconductor junctions in the presence of a spin-splitting
		magnetic field}.
	\newblock \emph{\bibinfo{journal}{Phys. Rev. Lett.}}
	\textbf{\bibinfo{volume}{112}}, \bibinfo{pages}{057001}
	(\bibinfo{year}{2014}).
	
	\bibitem{expcap1}
	\bibinfo{author}{Roddaro, S.} \emph{et~al.}
	\newblock \bibinfo{title}{Large thermal biasing of individual gated
		nanostructures}.
	\newblock \emph{\bibinfo{journal}{Nano Research}} \textbf{\bibinfo{volume}{7}},
	\bibinfo{pages}{579--587} (\bibinfo{year}{2015}).
	
	\bibitem{expcap2}
	\bibinfo{author}{Fornieri, A.}, \bibinfo{author}{Blanc, C.},
	\bibinfo{author}{Bosisio, R.}, \bibinfo{author}{D'Ambrosio, S.} \&
	\bibinfo{author}{Giazotto, F.}
	\newblock \bibinfo{title}{Nanoscale phase engineering of thermal transport with
		a josephson heat modulator}.
	\newblock \emph{\bibinfo{journal}{Nat Nano}} \textbf{\bibinfo{volume}{advance
			online publication}} (\bibinfo{year}{2015}).
	\newblock \bibinfo{note}{Letter}.
	
	\bibitem{expcap3}
	\bibinfo{author}{De~Franceschi, S.}, \bibinfo{author}{Kouwenhoven, L.},
	\bibinfo{author}{Schonenberger, C.} \& \bibinfo{author}{Wernsdorfer, W.}
	\newblock \bibinfo{title}{Hybrid superconductor-quantum dot devices}.
	\newblock \emph{\bibinfo{journal}{Nat Nano}} \textbf{\bibinfo{volume}{5}},
	\bibinfo{pages}{703--711} (\bibinfo{year}{2010}).
	
	\bibitem{expcap4}
	\bibinfo{author}{van Dam, Jorden A.}, \bibinfo{author}{Nazarov, Yuli V.},
	\bibinfo{author}{Bakkers, Erik P. A. M.} , \bibinfo{author}{De Franceschi, Silvano}\& \bibinfo{author}{Kouwenhoven, Leo P.}
	\newblock \bibinfo{title}{Supercurrent reversal in quantum dots}.
	\newblock \emph{\bibinfo{journal}{Nature}} \textbf{\bibinfo{volume}{442}},
	\bibinfo{pages}{667-670} (\bibinfo{year}{2006}).
	
	
	\bibitem{Yoni}
	\bibinfo{author}{Dubi, Y.} \& \bibinfo{author}{Di~Ventra, M.}
	\newblock \bibinfo{title}{\textit{Colloquium} : Heat flow and thermoelectricity
		in atomic and molecular junctions}.
	\newblock \emph{\bibinfo{journal}{Rev. Mod. Phys.}}
	\textbf{\bibinfo{volume}{83}}, \bibinfo{pages}{131--155}
	(\bibinfo{year}{2011}).
	
	\bibitem{AndreevReflection}
	\bibinfo{author}{Chang, L.-F.} \& \bibinfo{author}{Bagwell, P.~F.}
	\newblock \bibinfo{title}{Ballistic josephson-current flow through an
		asymmetric superconductor\char21{}normal-metal\char21{}superconductor
		junction}.
	\newblock \emph{\bibinfo{journal}{Phys. Rev. B}} \textbf{\bibinfo{volume}{49}},
	\bibinfo{pages}{15853--15863} (\bibinfo{year}{1994}).
	
	\bibitem{voltage}
	\bibinfo{author}{Panaitov, G.}, \bibinfo{author}{Ryazanov, V.},
	\bibinfo{author}{Ustinov, A.} \& \bibinfo{author}{Schmidt, V.}
	\newblock \bibinfo{title}{Thermoelectric ac josephson effect in \{SNS\}
		junctions}.
	\newblock \emph{\bibinfo{journal}{Physics Letters A}}
	\textbf{\bibinfo{volume}{100}}, \bibinfo{pages}{301 -- 303}
	(\bibinfo{year}{1984}).
	
	\bibitem{Anderson}
	\bibinfo{author}{Anderson, P.~W.}
	\newblock \bibinfo{title}{Localized magnetic states in metals}.
	\newblock \emph{\bibinfo{journal}{Phys. Rev.}} \textbf{\bibinfo{volume}{124}},
	\bibinfo{pages}{41--53} (\bibinfo{year}{1961}).
	
	
		\bibitem{rodero}
		\bibinfo{author}{A Martín-Rodero and A Levy Yeyati}
		\newblock \bibinfo{title}{The Andreev states of a superconducting quantum dot: mean field versus exact numerical results}.
		\newblock \emph{\bibinfo{journal}{Journal of Physics: Condensed Matter}} \textbf{\bibinfo{volume}{24}},
		\bibinfo{pages}{385303} (\bibinfo{year}{2012}).
	
	\bibitem{arovas}
	\bibinfo{author}{Rozhkov, A. V. and Arovas, Daniel P.}
	\newblock \bibinfo{title}{Josephson Coupling through a Magnetic Impurity}.
	\newblock \emph{\bibinfo{journal}{Phys. Rev. Lett.}} \textbf{\bibinfo{volume}{82}},
	\bibinfo{pages}{2788--2791} (\bibinfo{year}{1999}).


\bibitem{zonda}
\bibinfo{author}{Žonda, M.}, \bibinfo{author}{Pokorný, V.},
\bibinfo{author}{Janiš, V.} \& \bibinfo{author}{Novotný, T.}
\newblock \bibinfo{title}{Perturbation theory of a superconducting 0 - $\pi$ impurity quantum phase transition}.
\newblock \emph{\bibinfo{journal}{Scientific Reports}} \textbf{\bibinfo{volume}{5}},
\bibinfo{pages}{8821} (\bibinfo{year}{2015}).
	
	\bibitem{Kondo}
	\bibinfo{author}{Clerk, A.~A.} \& \bibinfo{author}{Ambegaokar, V.}
	\newblock \bibinfo{title}{Loss of $\ensuremath{\pi}$-junction behavior in an
		interacting impurity josephson junction}.
	\newblock \emph{\bibinfo{journal}{Phys. Rev. B}} \textbf{\bibinfo{volume}{61}},
	\bibinfo{pages}{9109--9112} (\bibinfo{year}{2000}).
	
	\bibitem{ABS}
	\bibinfo{author}{Meng, T.}, \bibinfo{author}{Florens, S.} \&
	\bibinfo{author}{Simon, P.}
	\newblock \bibinfo{title}{Self-consistent description of andreev bound states
		in josephson quantum dot devices}.
	\newblock \emph{\bibinfo{journal}{Phys. Rev. B}} \textbf{\bibinfo{volume}{79}},
	\bibinfo{pages}{224521} (\bibinfo{year}{2009}).
	
	\bibitem{Zhu}
	\bibinfo{author}{Zhu, Y.}, \bibinfo{author}{feng Sun, Q.} \&
	\bibinfo{author}{han Lin, T.}
	\newblock \bibinfo{title}{Andreev bound states and the pi-junction transition
		in a superconductor/quantum-dot/superconductor system}.
	\newblock \emph{\bibinfo{journal}{Journal of Physics: Condensed Matter}}
	\textbf{\bibinfo{volume}{13}}, \bibinfo{pages}{8783} (\bibinfo{year}{2001}).
	
	\bibitem{Ambegauker}
	\bibinfo{author}{Ambegaokar, V.} \& \bibinfo{author}{Baratoff, A.}
	\newblock \bibinfo{title}{Tunneling between superconductors}.
	\newblock \emph{\bibinfo{journal}{Phys. Rev. Lett.}}
	\textbf{\bibinfo{volume}{10}}, \bibinfo{pages}{486--489}
	(\bibinfo{year}{1963}).
	
	\bibitem{Meir}
	\bibinfo{author}{Meir, Y.} \& \bibinfo{author}{Wingreen, N.~S.}
	\newblock \bibinfo{title}{Landauer formula for the current through an
		interacting electron region}.
	\newblock \emph{\bibinfo{journal}{Phys. Rev. Lett.}}
	\textbf{\bibinfo{volume}{68}}, \bibinfo{pages}{2512--2515}
	(\bibinfo{year}{1992}).
	
	\bibitem{Langreth}
	\bibinfo{author}{Haug, H.}, \bibinfo{author}{Jauho, A.-P.} \&
	\bibinfo{author}{Cardona, M.}
	\newblock \emph{\bibinfo{title}{Quantum kinetics in transport and optics of
			semiconductors}}, vol.~\bibinfo{volume}{2} (\bibinfo{publisher}{Springer},
	\bibinfo{year}{2008}).
	
	\bibitem{Dyne1}
	\bibinfo{author}{Dynes, R.~C.}, \bibinfo{author}{Garno, J.~P.},
	\bibinfo{author}{Hertel, G.~B.} \& \bibinfo{author}{Orlando, T.~P.}
	\newblock \bibinfo{title}{Tunneling study of superconductivity near the
		metal-insulator transition}.
	\newblock \emph{\bibinfo{journal}{Phys. Rev. Lett.}}
	\textbf{\bibinfo{volume}{53}}, \bibinfo{pages}{2437--2440}
	(\bibinfo{year}{1984}).
	
	\bibitem{Dyne2}
	\bibinfo{author}{Mitrović, B.} \& \bibinfo{author}{Rozema, L.~A.}
	\newblock \bibinfo{title}{On the correct formula for the lifetime broadened
		superconducting density of states}.
	\newblock \emph{\bibinfo{journal}{Journal of Physics: Condensed Matter}}
	\textbf{\bibinfo{volume}{20}}, \bibinfo{pages}{015215}
	(\bibinfo{year}{2008}).
	
	
\end{thebibliography}

\section*{Acknowledgements}

We thank M. Di Ventra and O. Entin-Wohlman for illuminating discussions. YM acknowledges support from ISF under grants 11/11 and 292/15. FG acknowledges the European Research Council under the European Union's Seventh Framework Program (FP7/2007-2013)/ERC Grant agreement No.~615187-COMANCHE for partial financial support.

\end{document}